\documentclass[a4paper,dvips,12pt]{article}
\usepackage{axodraw}
\usepackage{array,tabularx,epsfig,graphicx,rotating}
\usepackage{ifthen}
\usepackage{fancybox}
\usepackage{amsfonts}

\newcommand{\beq}{\begin{equation}}
\newcommand{\eeq}{\end{equation}}

\newcommand{\ud}{\mathrm{d}}

\begin{document}

\begin{center}
{\large \bf On parameterizations of the Nordheim function }

\vspace{5mm} 
{ \bf B. B. Levchenko\footnote{levtchen@mail.desy.de}}\\

\vspace*{2mm}
{\small \it Institute of Nuclear Physics, Moscow State University, Moscow, 119992 Russia }\\

\end{center}

\abstract{\small 
\vspace*{-7mm}
\noindent\begin{quote}
 Several parameterizations of tabulated values of the Nordheim function $\vartheta (y)$ 
and its complementary function $t(y)$ are proposed. 
The function $\vartheta (y)$ plays essential role in description of field emission.
\end{quote}
}

\vspace*{5mm}

\noindent 1.{\ }Under  the influence of a strong electric field  $F$ (at field strength above 10$^7$  
V/cm) the surface of solids and liquids  emits electrons. 
The phenomenon, known now  as electron field emission, was discovered in 1897 by Robert Wood, 
who found that  a high voltage applied between a pointed cathode and an anode 
caused a current to flow \cite{wood}\footnote{  
After considerable  experimenting R. Wood "succeeded  
in finding a method of producing X-rays by what  appears to be a new form of cathode 
discharge, which manifests itself as a blue arc between two minute balls of platinum 
in a very high vacuum" \cite{wood}.}. 
Later, in 1928, Fowler and Nordheim \cite{FN}  proposed a theory of field emission
from metals, based on the newly created quantum mechanics,  as electrons 
tunneling through a potential barrier.

  In the framework of the Fowler-Nordheim theory,  the current density of field emission
electrons can be written in the following form \cite{FN},\cite{SB}-\cite{Gomer}

\beq
J\,=\, A \frac{F^2}{\varphi\cdot t^2 (y)}\exp \Big\{-B\frac{\ \varphi ^{3/2}}{F}\vartheta 
(y) \Big\}\, ,
\label{eq:1}
\eeq 
where  J is  the current density in A/cm$^2$, $F$ is electric field  at surface in V/cm, 
$\varphi$  is the work function  in eV. The field-independent constants A and  B 
  and the variable  $y$ are 
 
\beq
A=\frac{e^3}{8\pi h}\,=\, 1.5414\cdot 10^{-6}, \ \   B=\frac{8\pi
\sqrt{2m}}{3eh}\,=\,6.8309\cdot 10^7,
\ \  y= \frac{\sqrt{e^3 F}}{\varphi} = 3.7947\cdot 10^{-4} 
\frac{\sqrt{F}}{\varphi}
\label{eq:2*}
\eeq 
where $-e$ is the charge on the electron, $m$ is the electron mass
and $h$ is  Planck's constant. Numerical coefficients in Eq.(\ref{eq:2*}) corresponds to
resent values of the physical constants \cite{PDG}.

The Nordheim function $\vartheta (y)$ takes into account  a lowering of the potential 
barrier due to the image potential (the Shottky effect) and  its distinction from an 
idealized triangular shape. The function $t(y)$ in the denominator of Eq.(1) is defined as
\beq
t(y)\,=\,\vartheta (y)-(2y/3)(d\vartheta /dy).
\label{eq:5}
\eeq  
The expression of $\vartheta (y)$  (with a corrected definition of $k$ 
\cite{BKH}) is following

\beq
\vartheta (y)\,=\,\sqrt{\frac{1+\sqrt{1-y^2}}{2}}\cdot
\Big ( E(k^2)- \frac{y^2 K(k^2)}{1+\sqrt{1-y^2}}\Big ),
\label{eq:3}
\eeq 
where
\beq
E(k^2)\,=\,\int_0^{\frac{\pi}{2}} \sqrt{1-k^2\sin ^2 \beta} \ \ud \beta ,\ \ \ \ 
K(k^2)\,=\,\int_0^{\frac{\pi}{2}} \frac{\ud \beta}{\sqrt{1-k^2\sin ^2 \beta}},\ \
k^2\,=\,\frac{2\sqrt{1-y^2}}{1+\sqrt{1-y^2}}.
\label{eq:4}
\eeq 
$K(k^2)$ and $E(k^2)$ are the complete elliptic integrals of the first and second kind.

\vspace*{3mm}

\noindent 2.{\ } It follows from Eq.(\ref{eq:3}) that  $\vartheta (y)$ has a complicated
structure,
and so for a long time  only tabulated values of $\vartheta (y)$ 
and $t(y)$ \cite{BKH}-\cite{Sh}   were used. From our view there is a need to have a convenient
parameterization of $\vartheta (y)$  valid in  whole range of $y\in [0,1]$.

Figs. 1a and 1b shows the behavior of $\vartheta (y)$ and $t(y)$ with function values 
(points) taken from \cite{BKH}-\cite{Sh}.
The function $\vartheta (y)$ varies significantly with $y$ (i.e. with the field variation), 
however  $t(y)$ is quite close to unity at all values of $y$.
 
To fit $\vartheta (y)$ by elementary functions, several functional forms were tested. From them
were chosen power functions Eq.(6), Eq.(7) and  polynomial functions $P^{(n)}$ of different order 
n=3$\div$8,  Eq.(\ref{eq:8}).
The corresponding functions $t(y)$ were calculated with use of Eq.(\ref{eq:5})
{
\beq
\hspace*{-32mm} \vartheta (y) \,=\, p_0 + p_1 y^{p_2},\hspace*{21mm}  t(y) \,=\, p_0 + p_1(1-\frac{2}{3}p_2) 
y^{p_2} ,
\label{eq:6}
\eeq
\beq
\hspace*{-45mm}\vartheta (y)\,=\, p_0 + p_1 y + p_2 y^{3/2},\hspace*{9mm} t(y)\,=\,  p_0 +
\frac{1}{3} p_1\cdot y ,
\label{eq:7}
\eeq
 \beq
\hspace*{-35mm}\vartheta (y) \,=\, P^{(n)}(y) = {\sum_{i=0}^n} \ p_i y^i,
\hspace*{7mm}   t(y)\,=\,  {\sum_{i=0}^n} \ 
(1-\frac{2}{3}i) p_i\cdot y^i .
\label{eq:8}
\eeq 
}                       

The fitted parameters $p_i$  are listed in the Table. The parameters were obtained with use of the MathCad 
 program package \cite{MC}. Figs 1a, 1c and  2a  demonstrate a very good  quality of fit 
 provided by (\ref{eq:6})-(\ref{eq:8}). However, at large $y$ Eqs.(\ref{eq:6})-(\ref{eq:7}) 
 predict for $t(y)$  values slightly different (less 1$\%$) from the tabulated values. 
 Use of  Eq.\ref{eq:8} with $n$=6,8,... allows us to eliminate 
 this minor discrepancy (Fig. 2b).

\hspace*{10.0cm}{\bf Table }
\vspace*{2mm}
{\begin{center}
\begin{tabular}{|r|r|r|r|r|r|}
\hline
 i  & $p_i$, Eq.(\ref{eq:6}) &$p_i$, Eq.(\ref{eq:7})& $P^{(4)}$ & $P^{(6)}$  & $P^{(8)}$  \\
\hline
 0  & 1.003         &0.994  &1.001   &  1.00  &1.00	\\
 1  & -1.005        &0.310   &-0.064  &-0.027  &-0.015	 \\ 
 2  & 1.684         &-1.301 & -1.394 &-1.729  &-1.947	\\  
 3  &               &       &0.657   &1.803   &3.263	 \\  
 4  &               &       &-0.201  &-2.002  &-6.841	 \\ 
 5  &               &       &	     &1.318   &10.085	 \\    
 6  &               &       &	     &-0.363  &-9.193	 \\    
 7  &               &       &	     &	      &4.635	 \\    
 8  &               &       &	     &	      &-0.987	 \\	   
\hline

\end{tabular}
\end{center}}

\newpage

\hspace*{-25mm}
\begin{minipage}[h]{.47\textwidth}
\includegraphics[height=9.5cm,width=9.5cm]{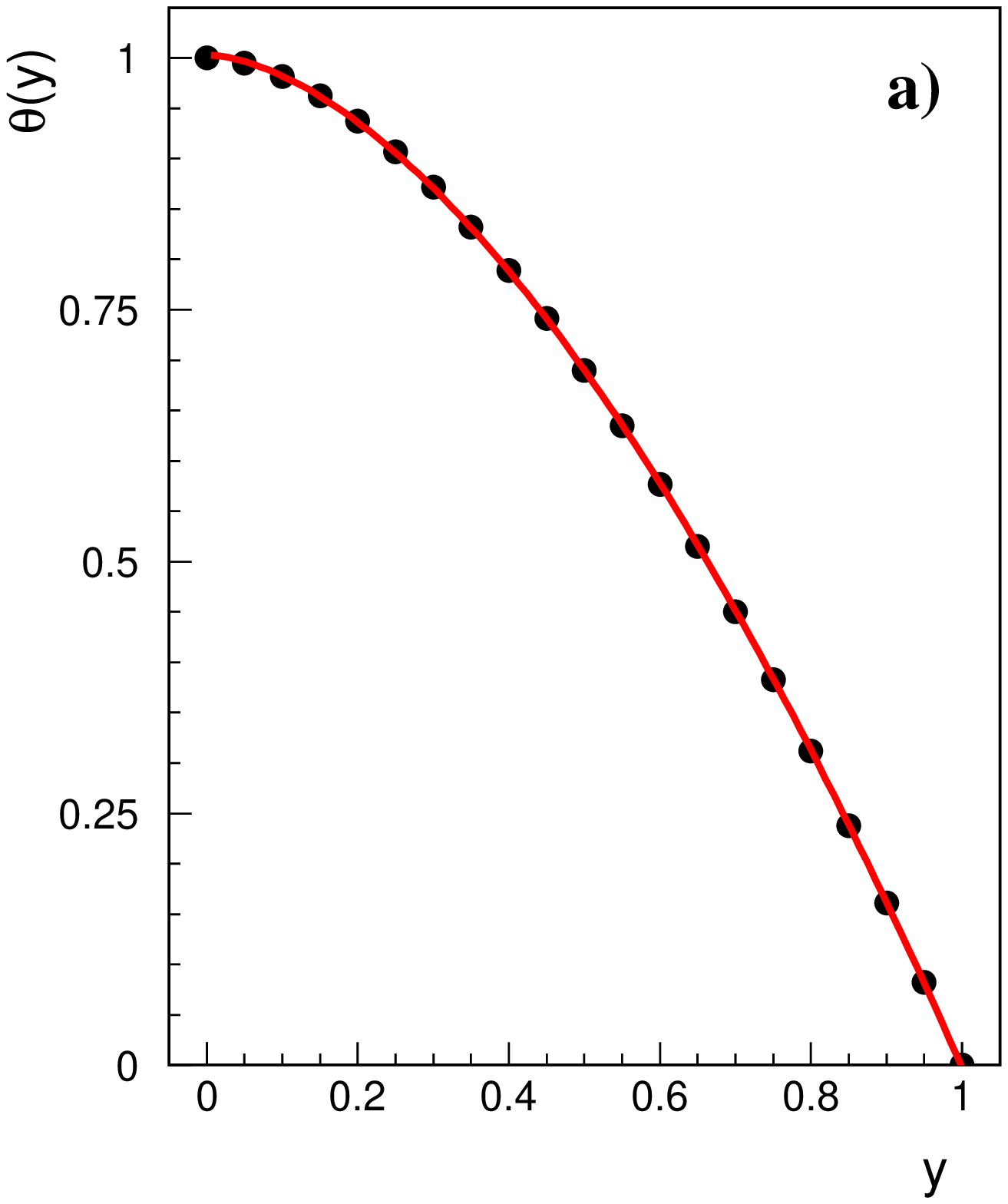}
\end{minipage}	\hspace*{+10mm}
\begin{minipage}[h]{.47\textwidth}
\includegraphics[height=9.5cm,width=9.5cm]{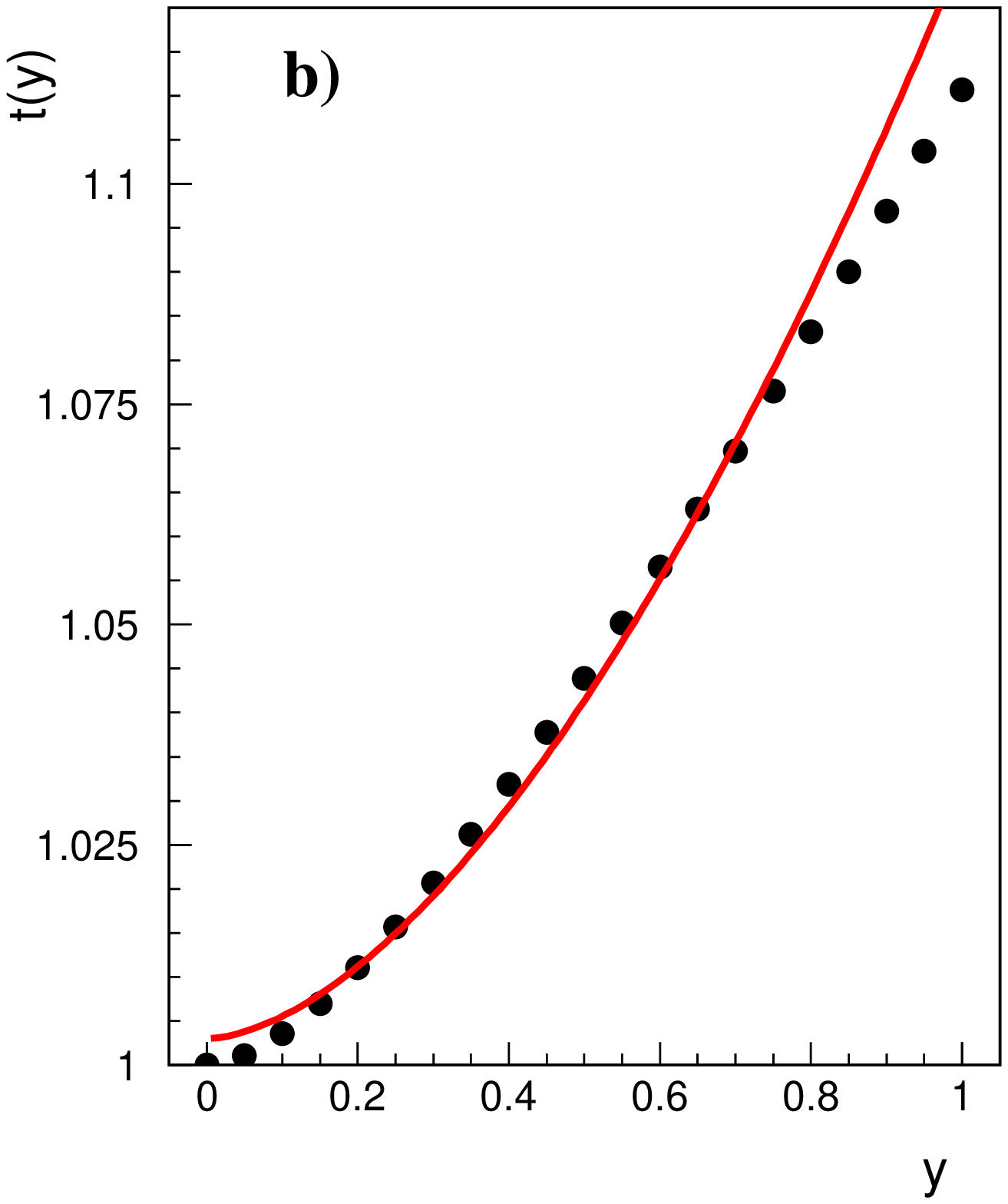}
\end{minipage}

\vspace*{-6mm}
\hspace*{-25mm}
\begin{minipage}[h]{.47\textwidth}
\includegraphics[height=9.5cm,width=9.5cm]{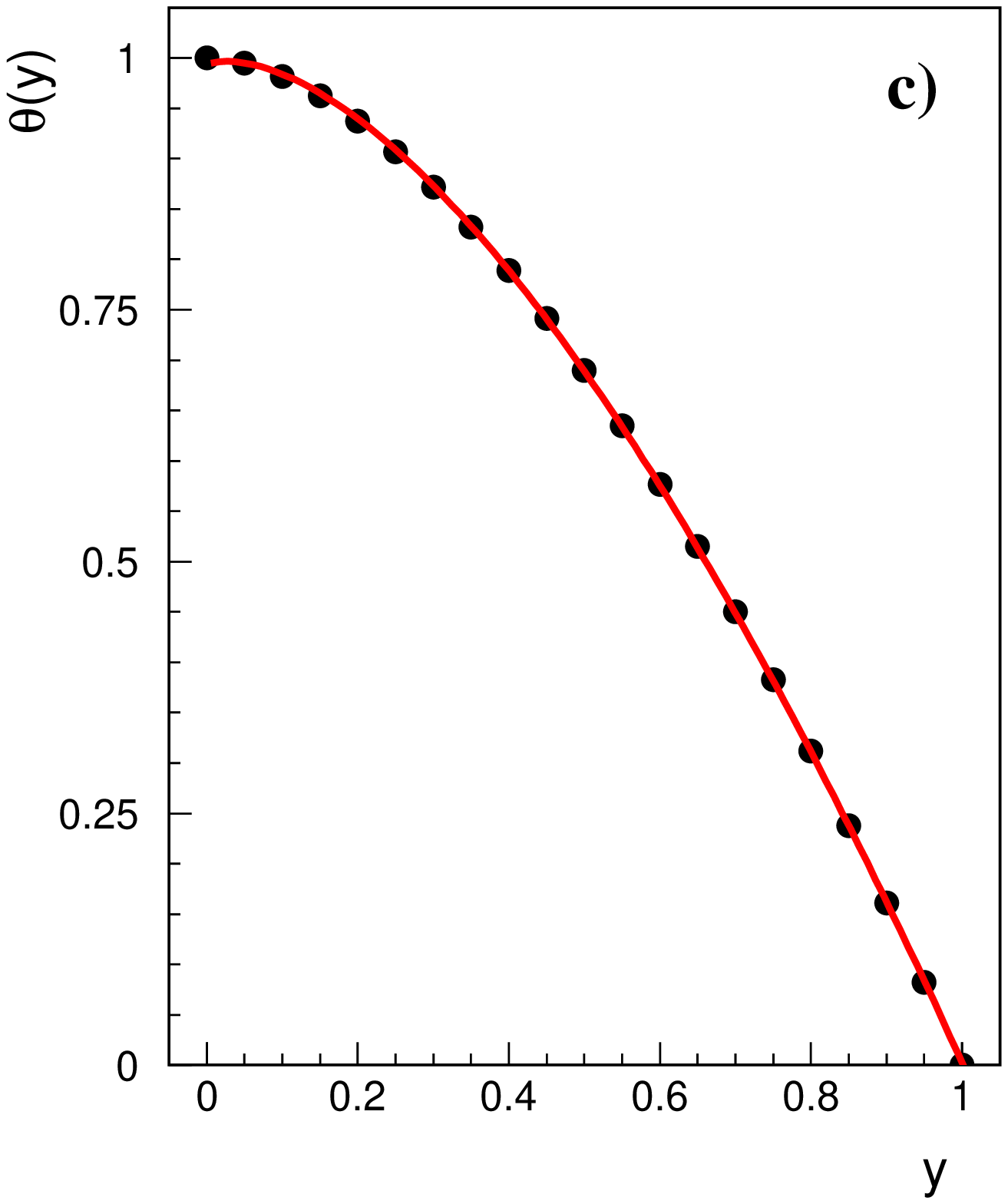}
\end{minipage}	\hspace*{+10mm}
\begin{minipage}[h]{.47\textwidth}
\includegraphics[height=9.5cm,width=9.5cm]{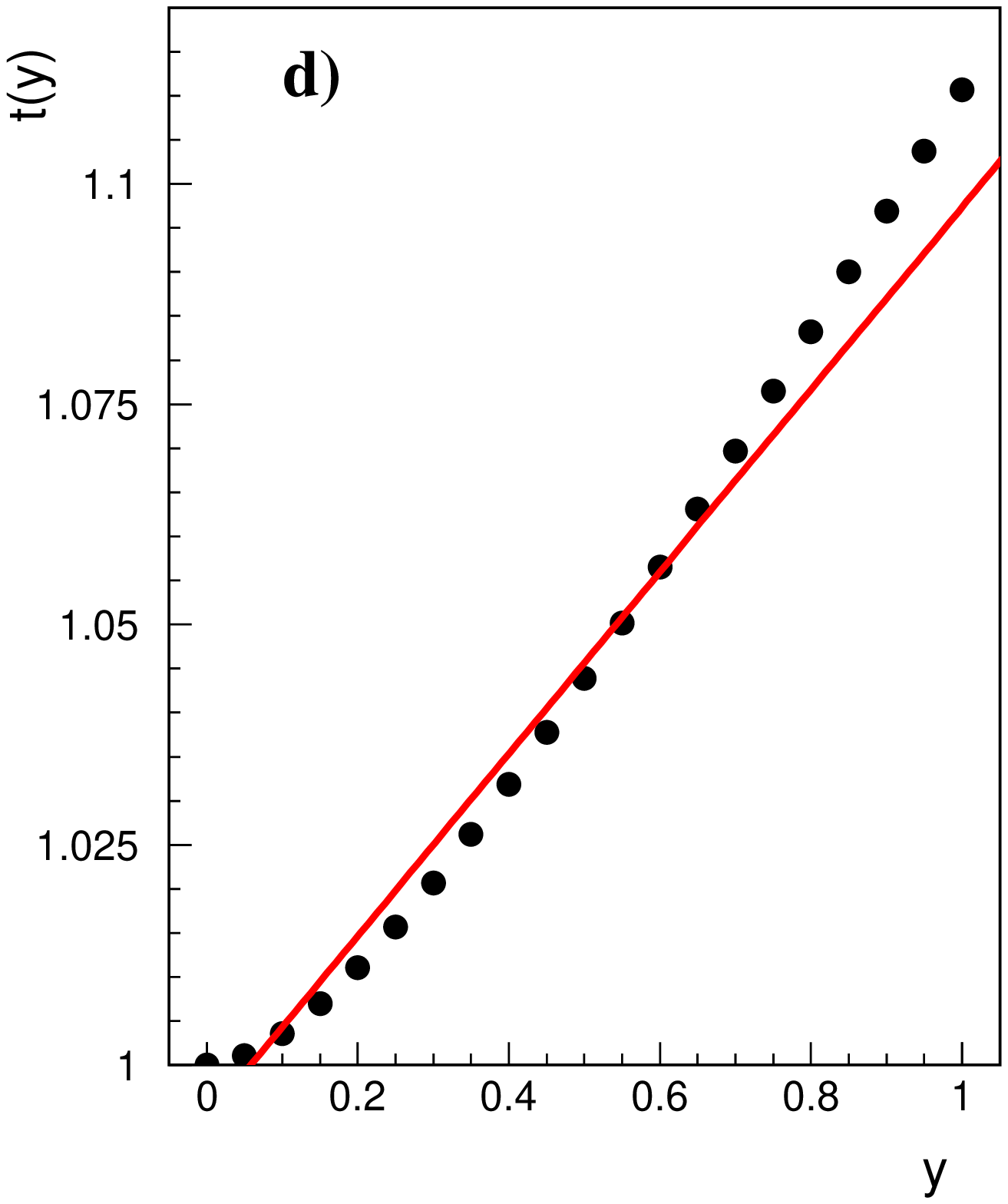}
\end{minipage}

\noindent
{\bf Figure~1:}
{\it  Functions $\vartheta (y)$ and $t(y)$ (points) parameterized (lines) by Eqs. (\ref{eq:6}),
 a), b)  and  Eqs. (\ref{eq:7}),  c), d).

 }

\newpage

\hspace*{-25mm}
\begin{minipage}[h]{.47\textwidth}
\includegraphics[height=9.5cm,width=9.5cm]{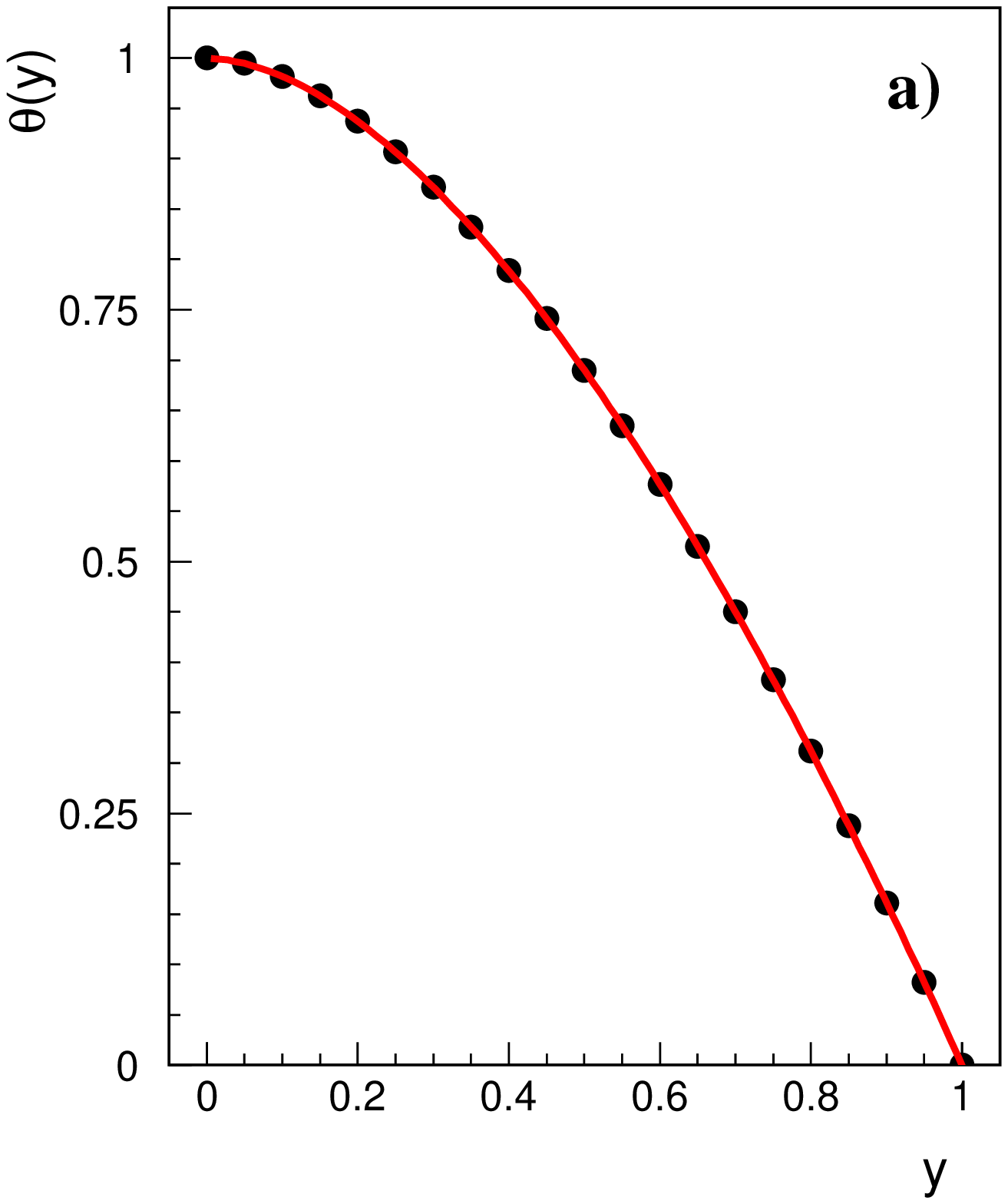}
\end{minipage}	\hspace*{+10mm}
\begin{minipage}[h]{.47\textwidth}
\includegraphics[height=9.5cm,width=9.5cm]{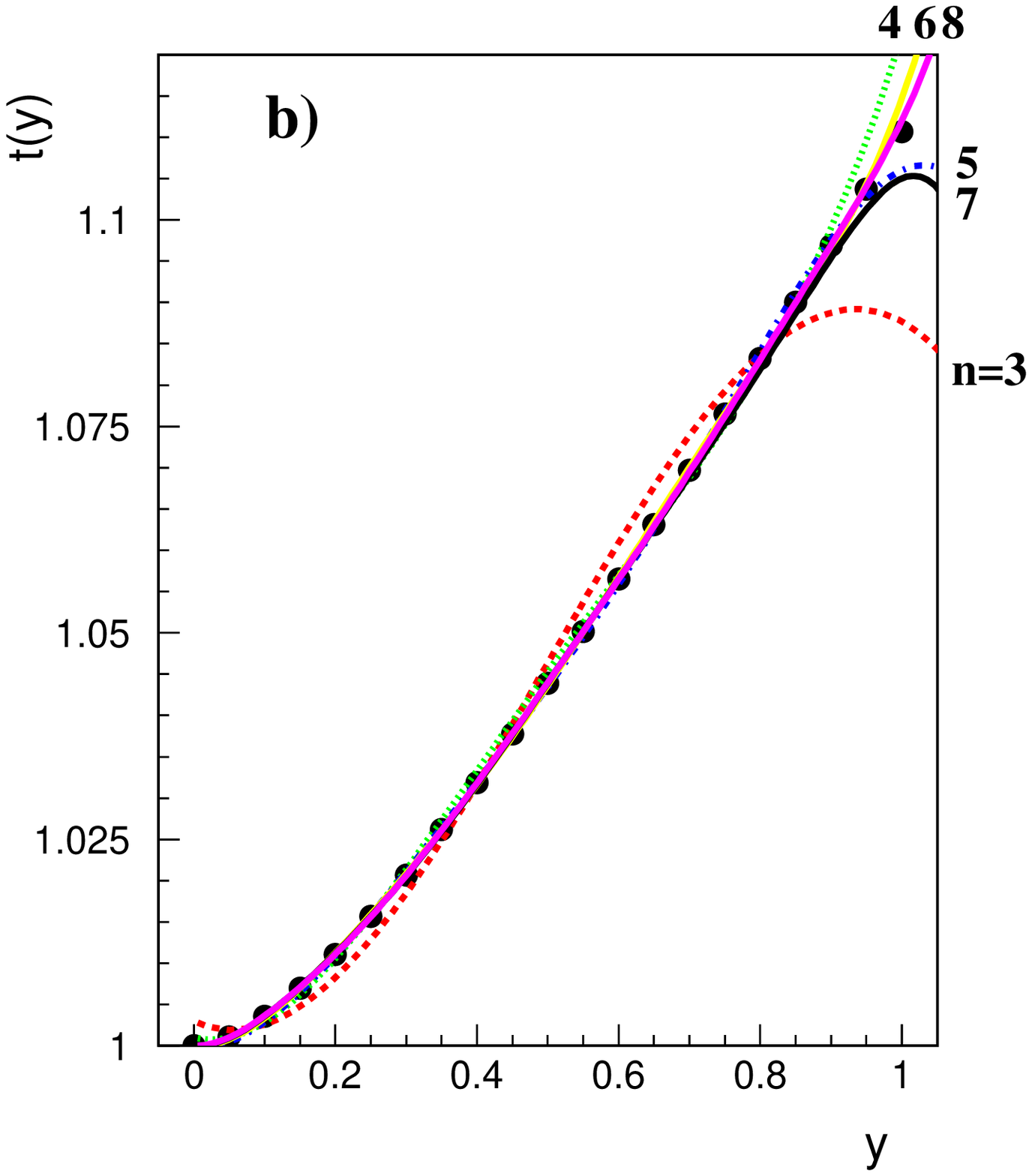}
\end{minipage}

\noindent
{\bf Figure~2:}{  \it Functions  $\vartheta (y)$ and  $t(y)$ (points) parameterized 
by Eqs. (\ref{eq:8}) (lines). 
 }
 
\vspace*{7mm}
To conclude,  the Nordheim elliptic function $\vartheta (y)$ can be well fitted
 by functions   of the type given by Eqs.(\ref{eq:6})-(\ref{eq:8}).

\vspace*{4mm}
\noindent
{ \it Acknowledgments.} The author is grateful to  E. Lohrmann and P. Bussey
for   useful discussions and comments.
This study is partially supported  by the Russian Foundation for Basic Research
under Grant no. 05-02-39028.

{}
\end{document}